\documentclass[conference,a4paper]{IEEEtran}
\IEEEoverridecommandlockouts
\usepackage{amsmath}
\usepackage{amsfonts}
\usepackage{amssymb}
\usepackage{algorithm}
\usepackage{algorithmic}
\usepackage{graphicx}
\usepackage{verbatim}

\newtheorem{theorem}{Theorem}
\newtheorem{lemma}{Lemma}

\newtheorem{corollary}{Corollary}
\newtheorem{example}{Example}

\newcommand{\beq}{\begin{equation}}
\newcommand{\eeq}{\end{equation}}
\newcommand{\beqnn}{\begin{equation*}}
\newcommand{\eeqnn}{\end{equation*}}
\newcommand{\beqy}{\begin{eqnarray}}
\newcommand{\eeqy}{\end{eqnarray}}
\newcommand{\beqynn}{\begin{eqnarray*}}
\newcommand{\eeqynn}{\end{eqnarray*}}
\newcommand{\bit}{\begin{itemize}}
\newcommand{\eit}{\end{itemize}}
\newcommand{\ben}{\begin{enumerate}}
\newcommand{\een}{\end{enumerate}}
\newcommand{\bex}{\begin{example}}
\newcommand{\eex}{\end{example}}

\newcommand{\balg}[1]{\begin{algorithm} \caption{#1}}
\newcommand{\ealg}{\end{algorithm}}

\newcommand{\balgc}{\begin{algorithmic}[1]}
\newcommand{\ealgc}{\end{algorithmic}}

\newcommand{\bary}{\begin{array}}
\newcommand{\eary}{\end{array}}
\newcommand{\bmx}{\begin{bmatrix}}
\newcommand{\emx}{\end{bmatrix}}
\newcommand{\bsmx}{\left[\begin{smallmatrix}}
\newcommand{\esmx}{\end{smallmatrix}\right]}
\newcommand{\bmxc}[1]{\left[\begin{array}{@{}#1@{}}}
\newcommand{\emxc}{\end{array}\right]}
\newcommand{\bcn}{\begin{center}}
\newcommand{\ecn}{\end{center}}

\newcommand{\A}{\boldsymbol{A}}

\newcommand{\I}{\boldsymbol{I}}

\renewcommand{\P}{\boldsymbol{P}}

\renewcommand{\S}{\boldsymbol{S}}

\newcommand{\rr}{\boldsymbol{r}}

\renewcommand{\v}{\boldsymbol{v}}

\newcommand{\x}{{\boldsymbol{x}}}
\newcommand{\y}{{\boldsymbol{y}}}

\newcommand{\0}{{\boldsymbol{0}}}

\usepackage{slashbox}
\usepackage{psfrag}
\usepackage{amssymb}
\usepackage{amsmath}
\usepackage{pifont}
\usepackage{cite}
\usepackage{graphics}
\usepackage{graphicx}
\usepackage{epsfig}
\usepackage{subfigure}
\usepackage{url}
\usepackage{amscd}
\usepackage{threeparttable}
\usepackage{enumerate}

\begin{document}

\title{A Sharp Condition for Exact Support Recovery of Sparse Signals With Orthogonal Matching Pursuit}
\author{\IEEEauthorblockN{
Jinming~Wen,
Zhengchun Zhou,
Jian Wang,
Xiaohu Tang and
Qun Mo}

%\\
%\IEEEauthorblockA{\IEEEauthorrefmark{1}College of Information Engineering and Shenzhen Key Lab of Media Security, Shenzhen University, \\
%Shenzhen, China, and ENS de Lyon, LIP,
%%(CNRS, ENS de Lyon, Inria, UCBL),
%Lyon 69007, France, Email: jwen@math.mcgill.ca}\\
%\IEEEauthorblockA{\IEEEauthorrefmark{2}School of Mathematics, Southwest Jiaotong University,
%Chengdu 610031, China, E-mail: zzc@home.swjtu.edu.cn}\\
%\IEEEauthorblockA{\IEEEauthorrefmark{3}Dept. of Electrical \& Computer Engineering, Seoul National University, Korea, E-mail: wangjianeee@gmail.com}\\
%\IEEEauthorblockA{\IEEEauthorrefmark{4}Information Security and National Computing Grid
%Laboratory, Southwest Jiaotong University,\\ Chengdu 610031, China, E-mail: xhutang@swjtu.edu.cn}\\
%\IEEEauthorblockA{\IEEEauthorrefmark{5}Department of Mathematics, Zhejiang University,
%Hangzhou 310027, China, E-mail: moqun@zju.edu.cn}

\thanks{J. Wen is with College of Information Engineering and Shenzhen Key Lab of Media Security, Shenzhen University, Shenzhen 518060, China. He is also with ENS de Lyon, LIP, Lyon 69007, France, (e-mail: jwen@math.mcgill.ca).}

\thanks{Z. Zhou is with School of Mathematics, Southwest Jiaotong University,
Chengdu 610031, China, (e-mail: zzc@home.swjtu.edu.cn).}

\thanks{J. Wang is with Dept. of Electrical \& Computer Engineering, Seoul National University, Seoul 151742, Korea, (e-mail: wangjianeee@gmail.com).}

\thanks{X. Tang is with Information Security and National Computing Grid
Laboratory, Southwest Jiaotong University, Chengdu 610031, China, (e-mail: xhutang@swjtu.edu.cn).}

\thanks{Q. Mo is with Dept. of Mathematics, Zhejiang University,
Hangzhou 310027, China, (e-mail: moqun@zju.edu.cn).}

\thanks{This work was supported by ``Programme Avenir
Lyon Saint-Etienne de l'Universit\'e de Lyon" in the framework of the programme
``Inverstissements d'Avenir" (ANR-11-IDEX-0007), ANR through the HPAC project under
Grant ANR~11~BS02~013, the NSF of China under grant 11271010, 11531013 and 61472024, and the fundamental research funds
for Central Universities.}}

\maketitle

\begin{abstract}
Support recovery of sparse signals from noisy measurements with orthogonal matching pursuit (OMP)
has been extensively studied in the literature.
In this paper, we show that for any $K$-sparse signal $\x$, if the sensing matrix $\A$ satisfies the restricted isometry property (RIP) of order $K + 1$ with restricted isometry constant (RIC)
$\delta_{K+1} < 1/\sqrt {K+1}$, then under some constraint on the minimum magnitude of the nonzero elements of $\x$, the OMP algorithm exactly recovers the support of $\x$ from the measurements $\y=\A\x+\v$ in $K$ iterations,
where $\v$ is the noise vector.
This condition is sharp in terms of $\delta_{K+1}$ since
for any given positive integer $K\geq 2$ and any $1/\sqrt{K+1}\leq t<1$,
there always exist a $K$-sparse $\x$ and a matrix $\A$ satisfying $\delta_{K+1}=t$
for which OMP may fail to recover the signal $\x$ in $K$ iterations.
Moreover, the constraint on the  minimum magnitude of the nonzero elements of $\x$ is weaker than existing results.
\end{abstract}

\begin{IEEEkeywords}
Compressed sensing (CS), restricted isometry property (RIP), orthogonal matching pursuit (OMP), support recovery.
\end{IEEEkeywords}

% \fi
\IEEEpeerreviewmaketitle

\section{Introduction}

In compressed sensing (CS), we usually observe the following linear model \cite{CanT05,Don06,CohDD09, WenLZ15}:
\beq
\label{e:model}
\y=\A\x+\v
\eeq
where $\x\in \mathbb{R}^n$ is an unknown $K$-sparse signal,
(i.e., $|\text{supp}(\x)|\leq K$, where $\text{supp}(\x)=\{i:x_i\neq0\}$ is the support of $\x$
and $|\text{supp}(\x)|$ is the cardinality of $\text{supp}(\x)$),
$\A\in \mathbb{R}^{m\times n}$ (with $m \ll n$) is a known sensing matrix,  $\v \in \mathbb{R}^{m}$ is a noise vector, and
$\y\in \mathbb{R}^m$ is the observation vector.
There are many types of noises, for example, the $l_2$ bounded noise ($\|\v\|_2\leq \epsilon$ for some constant $\epsilon$)
\cite{Fuc05,DonET06,Can08}, the $l_{\infty}$ bounded noise ($\|\A^T\v\|_{\infty}\leq \epsilon$) \cite{CaiW11},
and Gaussian noise ($v_i\sim \mathcal{N}(0, \sigma^2)$) \cite{CanT07}.
In this paper, we consider only the $l_2$ bounded noise.

One of the central goals of CS is to recover the signal $\x$ based on the sensing matrix $\A$ and measurement $\y$.
It has been revealed that under appropriate constraints on $\A$, reliable recovery of $\x$ can be achieved
 via properly designed algorithms (see, e.g., \cite{CanRT06,Mol11}).
%A common method to recover $\x$ from \eqref{e:model} is to solve the following $l_1-$minimization problem:
%\begin{eqnarray}
%\label{e:l1}
%\min_{\gamma\in \mathbb{R}^n}\|\gamma\|_1 :\; \;\text{subject \;to} \;
%\begin{cases}
%\|\y-\A\gamma\|_2\leq \epsilon, &\text{if}\;\epsilon\neq0
%\cr \y=\A\gamma , &\text{if}\;\epsilon=0
%\end{cases}.
%\end{eqnarray}
Orthogonal matching pursuit (OMP) \cite{TroG07} is a widely used algorithm for recovering sparse signals.
For any set $S\subset\{1,2,\ldots ,n\}$, let $\A_S$ be the submatrix of $\A$ that contains only the columns indexed by $S$,
and $\x_S$ be the subvector of $\x$ that contains only the entries indexed by $S$.
The OMP algorithm is described in Algorithm \ref{a:OMP}~\cite{TroG07}.

\begin{algorithm}[h!]
\caption{OMP}  \label{a:OMP}
Input: measurement $\y$, sensing matrix $\A$ and sparsity $K$.\\
Initialize: $k=0, \rr^0=\y, S_0=\emptyset$.\\
until stopping criterion is met
\begin{algorithmic}[1]
\STATE $k=k+1$,
\STATE $s^k=\arg \max\limits_{1\leq i\leq n}|\langle \rr^{k-1},\A_i\rangle|$,
\STATE $S_k=S_{k-1}\bigcup\{s^k\}$,
\STATE $\hat{\x}_{S_k}=\arg \min\|\y-\A_{S_k}\x\|_2$,
\STATE $\rr^k=\y-\A_{S_k}\hat{\x}_{S_k}$.
\end{algorithmic}
Output: $\hat{\x}=\arg \min\limits_{\x: \text{supp}(\x)=S_K}\|\y-\A\x\|_2$.
\end{algorithm}

A commonly used framework for analyzing CS recovery algorithms is the restricted isometry property (RIP) \cite{CanT05}.
For any $m\times n$ matrix $\A$ and any integer $K, 1\leq K\leq n$, the order $K$ restricted isometry constant (RIC) $\delta_K$ is
defined as the smallest constant such that
\begin{equation}
\label{e:RIP}
(1-\delta_K)\|\x\|_2^2\leq \|\A\x\|_2^2\leq(1+\delta_K)\|\x\|_2^2
\end{equation}
for all $K$-sparse vectors $\x$.
%If $k+k'\leq n$, then the $k,k'-$restricted orthogonality constant (ROC) $\theta_{k,k'}$ is defined as the smallest constant such that
%\beqnn
%|\langle\A\x, \A\x'\rangle|\leq \theta_{k,k'}\|\x\|_2\|\x'\|_2
%\eeqnn
%for all $\x$ and $\x'$, where $\x$ and $\x'$  are respectively $K$-sparse and $k'-$sparse and have disjoint supports.

%RIP conditions are hard to verify for a given matrix $\A$ \cite{CaiWX10}. So instead, one prefers to
%generate a matrix $\A$ randomly and show that the matrix satisfies the RIP with high probability \cite{BarDDW08} using the Johanson-Lindenstrauss lemma.

%Other frameworks in the CS analysis literature include the mutual incoherence property \cite{DonH01}.
%The mutual incoherence is defined by
%$$
%\mu=\max_{i\neq j}|\langle \A_i, \A_j\rangle|.
%$$

Many RIC-based conditions have been proposed to ensure exact recovery of sparse signals with OMP in the noise-free case.
It has respectively been shown in \cite{DavW10} and \cite{LiuT12} that $\delta_{K+1}<\frac{1}{3\sqrt{K}}$ and
$\delta_{K+1}<\frac{1}{(1+\sqrt{2})\sqrt{K}}$ are sufficient for OMP to recover any $K$-sparse $\x$ in $K$ iterations.
The condition has been improved to $\delta_{K+1}<\frac{1}{1+\sqrt{K}}$ in \cite{MoS12,WanS12},
and further improved to $\delta_{K+1}<\frac{\sqrt{4K+1}-1}{2K}$ in \cite{ChaW14}.
Recently, it is shown in \cite{Mo15} that if $\delta_{K+1}<\frac{1}{\sqrt{K+1}}$,
then OMP exactly recovers the $K$-sparse signal $\x$ in $K$ iterations.
On the other hand, it was conjectured in \cite{DaiM09} that there exist a matrix $\A$ with $\delta_{K+1}\leq\frac{1}{\sqrt{K}}$
and a $K$-sparse $\x$ such that OMP fails to recover $\x$ in $K$ iterations.
Examples provided in \cite{MoS12,WanS12} confirmed this conjecture.
Later, the example in \cite{WenZL13} showed that for any given positive integer $K\geq 2$ and
for any given $t$ satisfying $\frac{1}{\sqrt{K+1}}\leq t<1$, there always exist a $K$-sparse $\x$
and a matrix $\A$ satisfies the RIP of order $K+1$ with $\delta_{K+1}=t$
such that OMP may fail to recover the signal $\x$ in $K$ iterations.
In other words, the sufficient condition for recovering $\x$ cannot be weaker than $\delta_{K+1}<\frac{1}{\sqrt{K+1}}$.
Thus, $\delta_{K+1}<\frac{1}{\sqrt{K+1}}$ \cite{Mo15} is a sharp condition guaranteeing exact recovery of $K$-sparse signals with the OMP algorithm.

%Sufficient conditions based on the mutual incoherence and the minimum magnitude of the nonzero elements of the $K$-sparse signal $\x$ have been proposed to
%recover the support of $\x$ ($\text{supp}(\x)$) in the noise case have been proposed in \cite{CaiW11}.

For the noisy case, we are interested in recovering the support of $\x$,
since the signal can be estimated by an ordinary least squares regression on the recovered support \cite{CaiW11}.
%Under some conditions on the sensing matrix $\A$ and the minimum magnitude of the nonzero elements of the $K$-sparse signal $\x$,
%$\text{supp}(\x)$ can be exactly recovered by the OMP algorithm under the $l_{2}$ bounded noise.
It was shown in \cite{SheL15} that under some condition on the minimum magnitude of the nonzero elements of $\x$,
$\delta_{K+1} < \frac{1}{\sqrt {K}+3 }$ is sufficient for exact recovery of $\text{supp}(\x)$ with OMP under the $l_{2}$ bounded noise.
This condition has been improved to $\delta_{K+1} < \frac{1}{\sqrt {K}+1 }$ \cite{WuHC13}.
And the best existing condition in terms of $\delta_{K+1}$ is $\delta_{K+1}<\frac{\sqrt{4K+1}-1}{2K}$ \cite{ChaW14}.
%Note that the conditions on the minimum magnitude of the nonzero elements of $\x$ for the aforementioned conditions are different.
%The sufficient condition based on the mutual incoherence
%and the minimum magnitude of the nonzero elements of the $K$-sparse signal $\x$ has been proposed in \cite{CaiW11}.

In this paper, we investigate the RIP condition and the minimum magnitude of the nonzero elements of the $K$-sparse signal $\x$
that guarantee the recovery of $\text{supp}(\x)$ with OMP under the $l_{2}$ bounded noise ($\|\rr^k\|\leq \epsilon$). We show that if $\A$ and $\v$ in \eqref{e:model} respectively satisfy the RIP of order $K+1$ with
\begin{equation}
\delta_{K+1}<\frac{1}{\sqrt{K+1}}, \label{e:delta}
\end{equation}
then the OMP algorithm with stopping criterion $\|\rr^k\|\leq \epsilon$
exactly recovers $\text{supp}(\x)$ provided that
\begin{equation}
\min_{i\in\text{supp}(\x)}|x_i|>\frac{2\epsilon}{1-\sqrt {K+1 }\delta_{K+1}}. \label{e:xmin}
\end{equation}
By the aforementioned analysis, condition \eqref{e:delta} is sharp in terms of  $\delta_{K+1}$. We also show that condition \eqref{e:xmin} on $\min_{i\in\text{supp}(\x)}|x_i|$ is also weaker than existing results.

The rest of the paper is organized as follows.
%In section \ref{s:lemmas}, we give some useful lemmas which are needed to prove our main results.
In section \ref{s:main}, we present a sharp condition
for the exact support recovery of the $K$-sparse signal $\x$ by OMP under the $l_{2}$ bounded noise.
In section \ref{s:com}, we compare our sufficient condition with existing ones.
Finally,  we summarize this paper in section \ref{s:con}.

{\it Notation:} Let $\mathbb{R}$ be the real field. Boldface lowercase letters denote column vectors, and boldface uppercase letters denote matrices,
e.g., $\x\in\mathbb{R}^n$ and $\A\in\mathbb{R}^{m\times n}$.
%For a vector $\x$, $\x_{i:j}$ denotes the subvector of $\x$ formed by entries $i, i+1, \ldots,j$.
%Let $\e_k$ denote the $k$-th column of the identity matrix $\I$
%and $\e$ denote a column vector with all of its entries being 1,
%and $\0$ denote a zero matrix or a zero column vector.
Let $\Omega=\text{supp}(\x)$, then $|\Omega|\leq K$ for any $K$-sparse signal $\x$, where $|\Omega|$ is the cardinality of $\Omega$.
Let $\Omega \setminus S=\{k|k\in\Omega,k\not\in S\}$ for set $S$.
Let $\Omega^c$ and $S^c$ be the complement of $\Omega$ and $S$,
i.e., $\Omega^c=\{1,2,\ldots ,n\}\setminus \Omega$, and $S^c=\{1,2,\ldots ,n\}\setminus S$.
Let $\A_S$ be the submatrix of $\A$ that  contains only the columns indexed by $\S$,
and $\x_S$ be the subvector of $\x$ that  contains only the entries indexed by $\S$,
and $\A_S^T$ be the transpose of $\A_S$.
For full column rank matrix $\A_S$, let $\P_S=\A_S(\A_S^T\A_S)^{-1}\A_S^T$
and $\P^{\bot}_S=\I-\P_S$ denote the projector and orthogonal complement projector
on the column space of $\A_S$, respectively.

%$\e_k$ denotes the $k$-th column of the identity matrix $\I$.
%For $\x\in \mathbb{R}^n$, we use $\lfloor \x\rceil$ to denote its nearest integer vector, i.e.,
%each entry of $\x$ is rounded to its nearest integer (if there is a tie, the one with smaller magnitude is chosen).
%For a vector $\x$, $\x_{i:j}$ denotes the subvector of $\x$ formed by entries $i, i+1, \ldots,j$.
%For a matrix $\A$, $\A_{i:j,i:j}$ denotes the submatrix of $\A$ formed by rows and columns $i, i+1, \ldots,j$.
%The success probabilities of the Babai point and the ILS estimator are denoted by $P_{\sB}$ and $P_{\sILS}$, respectively.

\section{A sharp condition for exact support recovery under the $l_{2}$ bounded noise}
\label{s:main}

In this section, we show that if $\A$ satisfies the RIP of order $K+1$ with $\delta_{K+1} < \frac{1}{\sqrt {K+1 }}$,
then under some condition on the minimum magnitude of the nonzero elements of the $K$-sparse signal $\x$,
$\text{supp}(\x)$ can be exactly recovered by OMP under the $l_{2}$ bounded noise.

Before introducing our main result, we present the following lemma which is inspired by \cite{Mo15}.
\begin{lemma}
\label{l:main}
Suppose that $\A$ in \eqref{e:model} satisfies the RIP of order $K+1$ with $0\leq\delta_{K+1}<1$.
Let $S$ be a subset of $\Omega=\text{supp}(\x)$ with $|S|< |\Omega|$.
Then,
\begin{align}
\label{e:main}
\,\;&\|\A_{\Omega\setminus S}^T\P^{\bot}_S\A_{\Omega\setminus S}\x_{\Omega\setminus S}\|_{\infty}
-\|\A_{\Omega^c}^T\P^{\bot}_S\A_{\Omega\setminus S}\x_{\Omega\setminus S}\|_{\infty} \nonumber\\
\geq &\frac{(1-\sqrt {|\Omega|-|S|+1 }\delta_{|\Omega|+1})\|\x_{\Omega\setminus S}\|_2}{\sqrt{|\Omega|-|S|}}.
\end{align}
\end{lemma}

Due to the page limit, we skip the proof of Lemma \ref{l:main} and only give  an easily-checked example to explain the lemma.
Interested readers are referred to \cite{WenZWM15} for a detailed proof.

\textit{Example}:
Let $K=2$ and $S=\{1\}$. For $0\leq\delta<1$, let
\[
\A=
\bmx
\sqrt{1+\delta}&0&0\\
0&\sqrt{1-\delta}&0\\
0&0&\sqrt{1+\delta}
\emx~ \text{and}~
\x=
\bmx
1\\
1\\
0
\emx,
\]
then $\x$ is 2-sparse and $\Omega=\{1,2\}$. It is clear that
\[
\P^{\bot}_S=\P^{\bot}_{\{1\}}=
\bmx
0&0&0\\
0&1&0\\
0&0&1
\emx.
\]
Also, it is easily checked that $\delta_3=\delta$ and
\begin{align*}
\,\;&\|\A_{\Omega\setminus S}^T\P^{\bot}_S\A_{\Omega\setminus S}\x_{\Omega\setminus S}\|_{\infty}
-\|\A_{\Omega^c}^T\P^{\bot}_S\A_{\Omega\setminus S}\x_{\Omega\setminus S}\|_{\infty} \\
=&|\A_{\{2\}}^T\P^{\bot}_{\{1\}}\A_{\{2\}}\x_{\{2\}}|
-|\A_{\{3\}}^T\P^{\bot}_{\{1\}}\A_{\{2\}}\x_{\{2\}}|\\
=&1-\delta.
\end{align*}
One can show that
\[
\frac{(1-\sqrt {|\Omega|-|S|+1 }\delta_{|\Omega|+1})\|\x_{\Omega\setminus S}\|_2}{\sqrt{|\Omega|-|S|}}
=1-\sqrt{2}\delta.
\]
By the aforementioned two equations, \eqref{e:main} obviously holds in this case.

Since $|\Omega|\leq K$, from \eqref{e:main} it is not hard to see that under \eqref{e:delta}, the right-hand side of \eqref{e:main} is positive.

The following theorem gives a sufficient condition for exactly recovering $\text{supp}(\x)$ with OMP.

\begin{theorem}
\label{t:l2}
Suppose that $\A$ and $\v$ in \eqref{e:model} satisfy \eqref{e:delta}
and $\|\v\|_2\leq \epsilon$, respectively.
Then the OMP algorithm with stopping criterion $\|\rr^k\|\leq \epsilon$
exactly recovers the support $\Omega$ of the $K$-sparse signal $\x$ provided that
\beq
\label{e:l2}
\min_{i\in\Omega}|x_i|>\frac{2\epsilon}{1-\sqrt {K+1 }\delta_{K+1}}.
\eeq
\end{theorem}

Before proving Theorem \ref{t:l2}, we introduce three lemmas that are useful for our analysis.

\begin{lemma}[\cite{CanT05}]
\label{l:monot}
If $\A$ satisfies the RIP of orders $k_1$ and $k_2$ with $k_1<k_2$, then
$
\delta_{k_1}\leq \delta_{k_2}.
$
\end{lemma}

\begin{lemma}[\cite{NeeT09}]
\label{l:AtRIP}
Let $\A$ satisfy the RIP of order $k$ and $S$ be a set with $|S|\leq k$, then for any $\x \in \mathbb{R}^m$,
\beqnn
\|\A^T_S\x\|_2^2\leq(1+\delta_k)\|\x\|_2^2.
\eeqnn
\end{lemma}

\begin{lemma}[\cite{SheLPL14}]
\label{l:orthogonalcomp}
Let sets $S_1,S_2$ satisfy $|S_2\setminus S_1|\geq1$ and matrix $\A$ satisfy the RIP of order $|S_1\cup S_2|$, then
for any vector $\x \in \mathbb{R}^{|S_2\setminus S_1|}$,
\beqnn
(1-\delta_{|S_1\cup S_2|})\|\x\|_2^2\leq \|\P^{\bot}_{S_1}\A_{S_2\setminus S_1}\x\|_2^2\leq(1+\delta_{|S_1\cup S_2|})\|\x\|_2^2.
\eeqnn
\end{lemma}

{\em Proof of Theorem \ref{t:l2}}.  We prove the theorem in two steps. First, we show that OMP selects correct indexes in all iterations.
In the second step, we prove that the algorithm performs exactly $|\Omega|$ iterations before stopping.

We prove the first step by induction. Suppose that OMP selects correct indexes in the first $k-1$ iterations,
i.e., $S_{k-1}\subseteq \Omega$.
Then, we will show that the OMP algorithm also selects a correct index in the $k$-th iteration, that is, $s^k\in \Omega$.
Here, we assume $1\leq k\leq |\Omega|$, thus the proof for the first selection is contained in the case that $k=1$. Also, the induction assumption $S_{k-1}\subseteq \Omega$ holds in this case since $S_0=\emptyset$.

Obviously, for $i\in S_{k-1}$, $\langle \rr^{k-1},\A_i\rangle=0$. Thus
by line 2 of Algorithm \ref{a:OMP}, to show $s^k\in \Omega$, it suffices to show
\beq
\label{e:l2cond}
\max_{i\in \Omega\setminus S_{k-1}}|\langle \rr^{k-1},\A_i\rangle|
> \max_{j\in \Omega^c}|\langle \rr^{k-1},\A_j\rangle|.
\eeq

From line 4 of Algorithm \ref{a:OMP}, we have
\begin{align}
\label{e:xtk-1}
\hat{\x}_{S_{k-1}}=(\A_{S_{k-1}}^T\A_{S_{k-1}})^{-1}\A_{S_{k-1}}^T\y.
\end{align}
Thus, by line 5 of Algorithm \ref{a:OMP} and \eqref{e:xtk-1}, we have
\begin{align}
\label{e:rk-1}
\rr^{k-1}&=\y-\A_{S_{k-1}}\hat{\x}_{S_{k-1}}\nonumber \\
&=\big(\I-\A_{S_{k-1}}(\A_{S_{k-1}}^T\A_{S_{k-1}})^{-1}\A_{S_{k-1}}^T\big)\y \nonumber \\
&\overset{(a)}{=}\P^{\perp}_{S_{k-1}}(\A\x+\v) \nonumber \\
&\overset{(b)}{=}\P^{\perp}_{S_{k-1}}(\A_{\Omega}\x_{\Omega}+\v) \nonumber \\
&\overset{(c)}{=}\P^{\perp}_{S_{k-1}}(\A_{S_{k-1}}\x_{S_{k-1}}+\A_{\Omega\setminus S_{k-1}}\x_{\Omega\setminus S_{k-1}}+\v) \nonumber \\
&\overset{(d)}{=}\P^{\perp}_{S_{k-1}}\A_{\Omega\setminus S_{k-1}}\x_{\Omega\setminus S_{k-1}}+\P^{\perp}_{S_{k-1}}\v,
\end{align}
where (a), (b), (c) and (d) follow from the definition of $\P^{\perp}_{S_{k-1}}$,
the fact that $\Omega=\text{supp}(\x)$, the induction assumption $S_{k-1}\subseteq \Omega$, and
$\P^{\perp}_{S_{k-1}}\A_{S_{k-1}}=\0$, respectively.

Then it follows from \eqref{e:rk-1} that
\begin{align}
\label{e:left}
\,\;&\max_{i\in \Omega\setminus S_{k-1}}|\langle \rr^{k-1},\A_i\rangle| \nonumber\\
=&\|\A^T_{\Omega\setminus S_{k-1}}\big(\P^{\perp}_{S_{k-1}}\A_{\Omega\setminus S_{k-1}}\x_{\Omega\setminus S_{k-1}}+\P^{\perp}_{S_{k-1}}\v\big)\|_{\infty}
\nonumber\\
\geq&\|\A^T_{\Omega\setminus S_{k-1}}\P^{\perp}_{S_{k-1}}\A_{\Omega\setminus S_{k-1}}\x_{\Omega\setminus S_{k-1}}\|_{\infty}\nonumber\\
&-\|\A^T_{\Omega\setminus S_{k-1}}\P^{\perp}_{S_{k-1}}\v\|_{\infty},
\end{align}
and
\begin{align}
\label{e:right}
\,\;&\max_{j\in \Omega^c}|\langle \rr^{k-1},\A_j\rangle|\nonumber\\
=&\|\A^T_{\Omega^c}\big(\P^{\perp}_{S_{k-1}}\A_{\Omega\setminus S_{k-1}}\x_{\Omega\setminus S_{k-1}}+\P^{\perp}_{S_{k-1}}\v\big)\|_{\infty}
\nonumber\\
\leq&\|\A^T_{\Omega^c}\P^{\perp}_{S_{k-1}}\A_{\Omega\setminus S_{k-1}}\x_{\Omega\setminus S_{k-1}}\|_{\infty}
+\|\A^T_{\Omega^c}\P^{\perp}_{S_{k-1}}\v\|_{\infty}.
\end{align}

Therefore, from \eqref{e:left} and \eqref{e:right}, to show \eqref{e:l2cond}, it suffices to show
\begin{align}
\,\;&\|\A^T_{\Omega\setminus S_{k-1}}\P^{\perp}_{S_{k-1}}\A_{\Omega\setminus S_{k-1}}\x_{\Omega\setminus S_{k-1}}\|_{\infty}\nonumber\\
-&\|\A^T_{\Omega^c}\P^{\perp}_{S_{k-1}}\A_{\Omega\setminus S_{k-1}}\x_{\Omega\setminus S_{k-1}}\|_{\infty} \nonumber\\
>& \|\A^T_{\Omega\setminus S_{k-1}}\P^{\perp}_{S_{k-1}}\v\|_{\infty}+\|\A^T_{\Omega^c}\P^{\perp}_{S_{k-1}}\v\|_{\infty}
\label{e:l2cond1}.
\end{align}
By induction assumption $S_{k-1}\subseteq \Omega$, we have
\begin{align}
\label{e:suppxk-1}
|\text{supp}(\x_{\Omega\setminus S_{k-1}})|=|\Omega|+1-k.
\end{align}
Thus,
\begin{align}
\label{e:minxk-1}
\|\x_{\Omega\setminus S_{k-1}}\|_2&\geq \sqrt {|\Omega|+1-k }\min_{i\in\Omega\setminus S_{k-1}}|x_i| \nonumber\\
&\geq \sqrt {|\Omega|+1-k }\min_{i\in\Omega}|x_i|.
\end{align}

In the following, we give a lower bound on the left-hand side of \eqref{e:l2cond1}.
Since $S_{k-1}\subseteq \Omega$ and $|S_{k-1}|=k-1$, using Lemma \ref{l:main}, we have
\begin{align}
\,\;&\|\A^T_{\Omega\setminus S_{k-1}}\P^{\perp}_{S_{k-1}}\A_{\Omega\setminus S_{k-1}}\x_{\Omega\setminus S_{k-1}}\|_{\infty}\nonumber\\
-&\|\A^T_{\Omega^c}\P^{\perp}_{S_{k-1}}\A_{\Omega\setminus S_{k-1}}\x_{\Omega\setminus S_{k-1}}\|_{\infty} \nonumber\\
\geq&\frac{(1-\sqrt {|\Omega|-k+2 }\delta_{|\Omega|+1})\|\x_{\Omega\setminus S_{k-1}}\|_2}{\sqrt{|\Omega|+1-k}}\nonumber\\
\overset{(a)}{\geq}&\frac{(1-\sqrt {K+1 }\delta_{|\Omega|+1})\|\x_{\Omega\setminus S_{k-1}}\|_2}{\sqrt{|\Omega|+1-k}}\nonumber\\
\overset{(b)}{\geq}&\frac{(1-\sqrt {K+1 }\delta_{K+1})\|\x_{\Omega\setminus S_{k-1}}\|_2}{\sqrt{|\Omega|+1-k}}\nonumber\\
\overset{(c)}{\geq}&(1-\sqrt {K+1 }\delta_{K+1})\min_{i\in\Omega}|x_i|,
\label{e:l2cond1left}
\end{align}
where (a) is because $k\geq1$ and $\x$ is $K$-sparse (i.e., $|\Omega|\leq K$); (b) follows from Lemma \ref{l:monot};
and (c) follows from \eqref{e:delta} and \eqref{e:minxk-1}.

Next, we give an upper bound for the right-hand side of \eqref{e:l2cond1}. Clearly there exist $i_0\in\Omega\setminus S_{k-1}$ and $j_0\in\Omega^c$
such that
\begin{align}
\label{e:l2v1}
\|\A^T_{\Omega\setminus S_{k-1}}\P^{\perp}_{S_{k-1}}\v\|_{\infty}&=|\A^T_{i_0}\P^{\perp}_{S_{k-1}}\v|,\\
\label{e:l2v2}
\|\A^T_{\Omega^c}\P^{\perp}_{S_{k-1}}\v\|_{\infty}&=|\A^T_{j_0}\P^{\perp}_{S_{k-1}}\v|.
\end{align}
Therefore
\begin{align}
\label{e:l2cond1right}
\,&\|\A^T_{\Omega\setminus S_{k-1}}\P^{\perp}_{S_{k-1}}\v\|_{\infty}+\|\A^T_{\Omega^c}\P^{\perp}_{S_{k-1}}\v\|_{\infty}\nonumber \\
=&|\A^T_{i_0}\P^{\perp}_{S_{k-1}}\v|+|\A^T_{j_0}\P^{\perp}_{S_{k-1}}\v|\nonumber \\
=&\|\A^T_{i_0\cup j_0}\P^{\perp}_{S_{k-1}}\v\|_1\nonumber \\
\overset{(a)}{\leq}&\sqrt{2}\|\A^T_{i_0\cup j_0}\P^{\perp}_{S_{k-1}}\v\|_2\nonumber \\
\overset{(b)}{\leq}& \sqrt{2(1+\delta_{K+1})}\|\P^{\perp}_{S_{k-1}}\v\|_2\nonumber \\
\overset{(c)}{\leq}& \sqrt{2(1+\delta_{K+1})}\epsilon,
\end{align}
where (a) is because $\A^T_{i_0\cup j_0}\P^{\perp}_{S_{k-1}}\v$ is a $2\times1$ vector,
(b) follows from Lemma \ref{l:AtRIP}, and (c) is because
\beq
\label{e:orthcompv}
\|\P^{\perp}_{S_{k-1}}\v\|_2\leq\|\P^{\perp}_{S_{k-1}}\|_2\|\v\|_2\leq\|\v\|_2\leq\epsilon.
\eeq

Finally, from \eqref{e:l2cond1left} and \eqref{e:l2cond1right}, \eqref{e:l2cond1} (or equivalently \eqref{e:l2cond}) is guaranteed by
\begin{align*}
(1-\sqrt {K+1 }\delta_{K+1})\min_{i\in\Omega}|x_i|> \sqrt{2(1+\delta_{K+1})}\epsilon.
\end{align*}
from which we obtain \eqref{e:delta}.
%\beqnn
%\min_{i\in\Omega}|x_i|> \frac{\sqrt{2(1+\delta_{K+1})}\epsilon}{1-\sqrt {K+1 }\delta_{K+1}}.
%\eeqnn
%Thus, by Lemma \ref{l:cond}, \eqref{e:l2cond1} is guaranteed by \eqref{e:l2}.
Therefore, under \eqref{e:l2}, the OMP algorithm selects a correct index in each iteration.

Now we proceed to the second step of our proof. We show that the OMP algorithm performs exactly $|\Omega|$ iterations before stopping. That is, $ \|\rr^k\|_2>\epsilon$ for $1\leq k<|\Omega|$ and $ \|\rr^{|\Omega|}\|_2\leq\epsilon$.

Since the OMP algorithm selects a correct index in each iteration under \eqref{e:l2}, by \eqref{e:rk-1}, we have that for $1\leq k<|\Omega|$,
\begin{align}
\|\rr^k\|_2&= \|\P^{\perp}_{S_{k}}\A_{\Omega\setminus S_{k}}\x_{\Omega\setminus S_{k}}+\P^{\perp}_{S_{k}}\v\|_2\nonumber\\
&\geq \|\P^{\perp}_{S_{k}}\A_{\Omega\setminus S_{k}}\x_{\Omega\setminus S_{k}}\|_2-\|\P^{\perp}_{S_{k}}\v\|_2\nonumber\\
&\overset{(a)}{\geq}\|\P^{\perp}_{S_{k}}\A_{\Omega\setminus S_{k}}\x_{\Omega\setminus S_{k}}\|_2-\epsilon\nonumber\\
&\overset{(b)}{\geq}\sqrt{1-\delta_{|\Omega|}}\|\x_{\Omega\setminus\S_k}\|_2-\epsilon\nonumber\\
&\overset{(c)}{\geq}\sqrt{1-\delta_{K+1}}\sqrt{|\Omega|-k}\min_{i\in\Omega}|x_i|-\epsilon\nonumber\\
&\geq\sqrt{1-\delta_{K+1}}\min_{i\in\Omega}|x_i|-\epsilon,
\label{e:rklbd}
\end{align}
where (a) is from \eqref{e:orthcompv}; (b) is from Lemma \ref{l:orthogonalcomp};
and (c) follows from Lemma \ref{l:monot} and \eqref{e:minxk-1}.
Thus, if
\beq
\label{e:earlycondl2}
\min_{i\in\Omega}|x_i|> \frac{2\epsilon}{\sqrt {1-\delta_{K+1}}},
\eeq
then $\|\rr^k\|_2>\epsilon$ for each $1\leq k< \Omega$.

Furthermore, by noting that
\beqnn
%\label{e:deltaineq}
1-\sqrt {K+1 }\delta_{K+1}\leq1-\delta_{K+1}\leq\sqrt {1-\delta_{K+1}}.
\eeqnn
we have
\beq
\label{e:condl22}
\frac{2\epsilon}{1-\sqrt {K+1 }\delta_{K+1}}\geq \frac{2\epsilon}{\sqrt {1-\delta_{K+1}}}.
\eeq
This, together with \eqref{e:earlycondl2}, implies that if \eqref{e:l2} holds, $\|\rr^k\|_2>\epsilon$ for each $1\leq k< \Omega$. In other words, the OMP algorithm does not terminate before the $|\Omega|$-th iteration.

Similarly, by \eqref{e:rk-1},
\begin{align*}
%\label{e:rKbd}
\|\rr^{|\Omega|}\|_2&= \|\P^{\perp}_{S_{|\Omega|}}\A_{\Omega\setminus S_{|\Omega|}}
\x_{\Omega\setminus S_{|\Omega|}}+\P^{\perp}_{S_{|\Omega|}}\v\|_2\nonumber\\
&\overset{(a)}{=}\|\P^{\perp}_{S_{|\Omega|}}\v\|_2
\overset{(b)}{\leq}\epsilon,
\end{align*}
where (a) is because $S_{|\Omega|}=|\Omega|$ and (b) follows from \eqref{e:orthcompv}.
Therefore, under stopping condition $\|\rr^k\|_2>\epsilon$, the OMP algorithm performs $|\Omega|$ iterations before stopping. This completes the proof.\ \ $\Box$

From Theorem \ref{t:l2}, if $\epsilon=0$, then $\|\v\|_2= 0$ and \eqref{e:l2} holds. Hence, $\text{supp}(\x)$ can be exactly recovered in $|\text{supp}(\x)|$ iterations
if $\delta_{K+1}$ satisfies \eqref{e:delta}.
We thus have the following result, which is equivalent to \cite[Theorem III.1]{Mo15}.
\begin{corollary}
\label{c:l2}
Suppose that $\A$ and $\v$ in \eqref{e:model} satisfy the RIP of order $K+1$ with $\delta_{K+1}$ satisfying \eqref{e:delta}
and $\|\v\|_2= 0$, respectively. Then the OMP algorithm exactly recovers the $K$-sparse signal $\x$ in $K$ iterations.
\end{corollary}

The example in \cite{WenZL13} showed that for any given positive integer $K\geq 2$ and for any $\frac{1}{\sqrt{K+1}}\leq t<1$,
there always exist a $K$-sparse $\x$ and a matrix $\A$ satisfying the RIP of order $K+1$ with $\delta_{K+1}=t$
such that the OMP algorithm may fail to recover $\x$.
Thus, the sufficient condition, given in Theorem \ref{t:l2},
is sharp in terms of $\delta_{K+1}$  for guaranteeing exact recovery of $\text{supp}(\x)$.

\section{Comparison with exiting sufficient conditions}
\label{s:com}

In this section, we show that our sufficient condition given in Theorem \ref{t:l2} is weaker than existing sufficient conditions.

In \cite{WuHC13,ChaW14}, $\A$ was assumed to be column normalized,
i.e., $\|\A_i\|_2=1$ for $i=1,2,\ldots, n$. Note that Theorem \ref{t:l2} obviously holds if $\A$ is column normalized.
In fact, our result in Theorem \ref{t:l2} outperforms those in \cite{SheL15,WuHC13,ChaW14}
in terms of both $\delta_{K+1}$ and the requirement on $\min_{i\in\Omega}|x_i|$.
For simplicity, we only compare our condition with the so far best result \cite{ChaW14}.
%Note that, although \cite{SheL15} is published in 2015, the results were proposed in 2013, which are earlier than that in \cite{ChaW14}.

It was shown in \cite{ChaW14} that if $\A$ in \eqref{e:model} is column normalized and satisfies the RIP of order $K+1$ with
$\delta_{K+1}$ satisfying
$$
\delta_{K+1}<\frac{\sqrt{4K+1}-1}{2K}
$$
and $\v$ in \eqref{e:model} satisfies $\|\v\|_2\leq \epsilon$.
Then the OMP algorithm with stopping criterion $\|\rr^k\|\leq \epsilon$  exactly recovers the support $\Omega$ of the $K$-sparse signal $\x$ if
$$
\min_{i\in\Omega}|x_i|>\frac{(\sqrt{1+\delta_{K+1}}+1)\epsilon}{1-\delta_{K+1}-\sqrt{1-\delta_{K+1}}\sqrt{K}\delta_{K+1}}.
$$

By Theorem \ref{t:l2}, to show our condition is better (weaker), we only need to show that
\beq
\label{e:compl2ric}
\frac{\sqrt{4K+1}-1}{2K}<\frac{1}{\sqrt{K+1}}
\eeq
and that
\beq
\label{e:compl2x}
\frac{(\sqrt{1+\delta_{K+1}}+1)\epsilon}{1-\delta_{K+1}-\sqrt{1-\delta_{K+1}}\sqrt{K}\delta_{K+1}}
\geq\frac{2\epsilon}{1-\sqrt {K+1 }\delta_{K+1}}
\eeq
for $\delta_{K+1}$ satisfying \eqref{e:delta}.
In particular, if $\delta_{K+1}\neq 0$, then the strict inequality in \eqref{e:compl2x} holds.

Clearly to show \eqref{e:compl2ric}, it suffices to show
$$
\sqrt{(4K+1)(K+1)}<2K+\sqrt {K+1 }.
$$
Equivalently,
$$
4K^2+5K+1<4K^2+K+1+4K\sqrt {K+1 }.
$$
In fact, since $K\geq1$, the above equation holds trivially, and hence \eqref{e:compl2ric} is true.

Next, we assume  $\delta_{K+1}\neq 0$ satisfies \eqref{e:delta} and then show the strict inequality in \eqref{e:compl2x} holds.
Since $\delta_{K+1}\neq 0$,
$$
\sqrt{1+\delta_{K+1}}+1>2.
$$
Thus, it suffices to show
$$
1-\delta_{K+1}-\sqrt{1-\delta_{K+1}}\sqrt{K}\delta_{K+1}<1-\sqrt {K+1 }\delta_{K+1},
$$
or equivalently,
\beq
\label{e:compl2x1}
1+\sqrt{1-\delta_{K+1}}\sqrt{K}>\sqrt{K+1}.
\eeq
Obviously, \eqref{e:compl2x1} holds if
$$
\sqrt{1-\delta_{K+1}}>\frac{\sqrt{K+1}-1}{\sqrt{K}},
$$
which is equivalent to
$$
\delta_{K+1}<\frac{2(\sqrt{K+1}-1)}{K}.
$$
Thus, a sufficient condition of \eqref{e:compl2x} is
$$
\frac{1}{\sqrt{K+1}}<\frac{2(\sqrt{K+1}-1)}{K}.
$$
By some simple calculations, one can easily show that the aforementioned inequality holds.
Therefore, the strict inequality in \eqref{e:compl2x} holds if $\delta_{K+1}\neq0$ satisfies \eqref{e:delta}.

\section{Conclusion}
\label{s:con}
In this paper, we have studied the condition for exact support recovery of sparse signals from noisy measurements with OMP.
We have shown that if the sensing matrix $\A$ satisfies
$\delta_{K+1} < \frac{1}{\sqrt {K+1} }$, then under some constraint on the minimum magnitude of the nonzero elements of the $K$-sparse signal $\x$,
the support of the signal can be exactly recovered under the $l_2$ bounded noise.
This condition is sharp in terms of $\delta_{K+1}$
and also the constraint on the minimum magnitude of the nonzero elements of $\x$ is weaker than existing ones.

\appendices

\bibliographystyle{IEEEtran}
\bibliography{ref-RIP}

\begin{thebibliography}{10}
\providecommand{\url}[1]{#1}
\csname url@samestyle\endcsname
\providecommand{\newblock}{\relax}
\providecommand{\bibinfo}[2]{#2}
\providecommand{\BIBentrySTDinterwordspacing}{\spaceskip=0pt\relax}
\providecommand{\BIBentryALTinterwordstretchfactor}{4}
\providecommand{\BIBentryALTinterwordspacing}{\spaceskip=\fontdimen2\font plus
\BIBentryALTinterwordstretchfactor\fontdimen3\font minus
  \fontdimen4\font\relax}
\providecommand{\BIBforeignlanguage}[2]{{%
\expandafter\ifx\csname l@#1\endcsname\relax
\typeout{** WARNING: IEEEtran.bst: No hyphenation pattern has been}%
\typeout{** loaded for the language `#1'. Using the pattern for}%
\typeout{** the default language instead.}%
\else
\language=\csname l@#1\endcsname
\fi
#2}}
\providecommand{\BIBdecl}{\relax}
\BIBdecl

\bibitem{CanT05}
E.~J. Cand{\'e}s and T.~Tao, ``Decoding by linear programming,'' \emph{IEEE
  Trans. Inf. Theory}, vol.~51, no.~12, pp. 4203--4215, 2005.

\bibitem{Don06}
D.~L. Donoho, ``Compressed sensing,'' \emph{IEEE Trans. Inf. Theory}, vol.~52,
  no.~4, pp. 1289--1306, 2006.

\bibitem{CohDD09}
A.~Cohen, W.~Dahmen, and R.~DeVore, ``Compressed sensing and best $k$-term
  approximation,'' \emph{J. Amer. Math. Soc.}, vol.~22, pp. 211--231, 2009.

\bibitem{WenLZ15}
J.~Wen, D.~Li, and F.~Zhu, ``Stable recovery of sparse signals via
  $l_p$-minimization,'' \emph{Applied and Computational Harmonic Analysis},
  vol.~38, no.~1, pp. 161--176, 2015.

\bibitem{Fuc05}
J.~J. Fuchs., ``Recovery of exact sparse representations in the presence of
  bounded noise,'' \emph{IEEE Trans. Inf. Theory}, vol.~51, no.~10, pp.
  3601--3608, 2005.

\bibitem{DonET06}
D.~L. Donoho, M.~Elad, and V.~N. Temlyakov, ``Stable recovery of sparse
  overcomplete representations in the presence of noise,'' \emph{IEEE Trans.
  Inf. Theory}, vol.~52, pp. 6--18, 2006.

\bibitem{Can08}
E.~J. Cand{\'e}s, ``The restricted isometry property and its implications for
  compressed sensing,'' \emph{C. R. Acad. Sci. Paris, Ser. I}, vol. 346,
  no.~11, pp. 589--592, 2008.

\bibitem{CaiW11}
T.~Cai and L.~Wang, ``Orthogonal matching pursuit for sparse signal recovery
  with noise,'' \emph{IEEE Trans. Inf. Theory}, vol.~57, pp. 4680--4688, 2011.

\bibitem{CanT07}
E.~J. Cand{\'e}s and T.~Tao, ``The dantzig selector: Statistical estimation
  when $p$ is much larger than $n$,'' \emph{Ann. Statist}, vol.~35, pp.
  2313--2351, 2007.

\bibitem{CanRT06}
E.~J. Cand{\'e}s, J.~Romberg, and T.~Tao, ``Stable signal recovery from
  incomplete and inaccurate measurements,'' \emph{Comm. Pure Appl. Math},
  vol.~59, pp. 1207--1223, 2006.

\bibitem{Mol11}
Q.~Mo and S.~Li, ``New bounds on the restricted isometry constant
  $\delta_{2k}$,'' \emph{Appl. Comput. Harmon. Anal.}, vol.~31, pp. 460--468,
  2011.

\bibitem{TroG07}
J.~A. Tropp and A.~C. Gilbert, ``Signal recovery from random measurements via
  orthogonal matching pursuit,'' \emph{IEEE Trans. Inf. Theory}, vol.~53,
  no.~12, pp. 4655--4666, 2007.

\bibitem{DavW10}
M.~Davenport and M.~Wakin, ``Analysis of orthogonal matching pursuit using the
  restricted isometry property,'' \emph{IEEE Trans. Inf. Theory}, vol.~56,
  no.~9, pp. 4395--4401, 2010.

\bibitem{LiuT12}
E.~Liu and V.~Temlyakov, ``The orthogonal super greedy algorithm and
  applications in compressed sensing,'' \emph{IEEE Trans. Inf. Theory},
  vol.~58, no.~4, pp. 2040--2047, 2012.

\bibitem{MoS12}
Q.~Mo and S.~Yi, ``A remark on the restricted isometry property in orthogonal
  matching pursuit,'' \emph{IEEE Trans. Inf. Theory}, vol.~58, no.~6, pp.
  3654--3656, 2012.

\bibitem{WanS12}
J.~Wang and B.~Shim, ``On the recovery limit of sparse signals using orthogonal
  matching pursuit,'' \emph{IEEE Trans. Signal Process.}, vol.~60, no.~9, pp.
  4973--4976, 2012.

\bibitem{ChaW14}
L.-H. Chang and J.-Y. Wu, ``An improved {RIP}-based performance guarantee for
  sparse signal recovery via orthogonal matching pursuit,'' \emph{IEEE Trans.
  Inf. Theory}, vol.~60, no.~9, pp. 707--710, 2014.

\bibitem{Mo15}
Q.~Mo, ``A sharp restricted isometry constant bound of orthogonal matching
  pursuit,'' \emph{arXiv:1501.01708}.

\bibitem{DaiM09}
M.~Davenport and M.~Wakin, ``Subspace pursuit for compressive sensing signal
  reconstruction,'' \emph{IEEE Trans. Inf. Theory}, vol.~55, no.~5, pp.
  2230--2249, 2009.

\bibitem{WenZL13}
J.~Wen, X.~Zhu, and D.~Li, ``Improved bounds on the restricted isometry
  constant for orthogonal matching pursuit,'' \emph{Electronics Letters},
  vol.~49, pp. 1487--1489, 2013.

\bibitem{SheL15}
Y.~Shen and S.~Li, ``Sparse signals recovery from noisy measurements by
  orthogonal matching pursuit,'' \emph{Inverse Problems and Imaging}, vol.~9,
  no.~1, pp. 231--238, 2015.

\bibitem{WuHC13}
R.~Wu, W.~Huang, and D.~Chen, ``The exact support recovery of sparse signals
  with noise via orthogonal matching pursuit,'' \emph{IEEE Signal Processing
  Letters}, vol.~20, no.~4, pp. 403--406, 2013.

\bibitem{WenZWM15}
J.~Wen, Z.~Zhou, J.~Wang, X.~Tang, and Q.~Mo, ``Sharp conditions for exact
  support recovery of sparse signals with noise via {OMP},'' \emph{arXiv
  preprint arXiv:1512.07248}, 2015.

\bibitem{NeeT09}
D.~Needel and J.~A. Tropp, ``Co{S}a{MP}: Iterative signal recovery from
  incomplete and inaccurate samples,'' \emph{Applied and Computational Harmonic
  Analysis}, vol.~26, no.~3, pp. 301--321, 2009.

\bibitem{SheLPL14}
Y.~Shen, B.~Li, W.~Pan, and J.~Li, ``Analysis of generalized orthogonal
  matching pursuit using restricted isometry constant,'' \emph{Electron.
  Lett.}, vol.~50, no.~14, pp. 1020--1022, 2014.

\end{thebibliography}
\end{document}